\documentclass[a4paper,fleqn,usenatbib]{mnras}

\usepackage[T1]{fontenc}
\usepackage{ae,aecompl}

\usepackage{graphicx}	
\usepackage{amsmath}	
\usepackage{amssymb}	
\usepackage{color}
\usepackage{ulem}
\usepackage{censor}

\newlength\nextcharwidth
\makeatletter
\renewcommand\@cenword[1]{%
  \setlength{\nextcharwidth}{\widthof{#1}}%
  \censorrule{\nextcharwidth}%
  \kern -\nextcharwidth%
  #1}
\makeatother

\title[Spiral structures in a quiescent accretion disc]{Spiral structures and 
temperature distribution in the quiescent accretion disc of the cataclysmic binary V2051~Ophiuchi}

\author[A. Rutkowski et al.]{
A. Rutkowski,$^{1}$\thanks{E-mail: rudy@camk.edu.pl}
W. Waniak,$^{1}$
G. Preston$^{2}$
and W. Pych${^3}$
\\
$^{1}$ Astronomical Observatory of the Jagiellonian University, ul. Orla 171, 30-244 Krak\'ow, Poland\\
$^{2}$Carnegie Observatories, 813 Santa Barbara Street, Pasadena, CA 78712, USA\\
$^{3}$Nicolaus Copernicus Astronomical Center, ul. Bartycka 18, 00-716 Warszawa, Poland
}

\date{Accepted XXX. Received YYY; in original form ZZZ}

\pubyear{2015}

\begin{document}
\label{firstpage}
\pagerange{\pageref{firstpage}--\pageref{lastpage}}
\maketitle

\begin{abstract}
We present the capabilities of our new code for obtaining Doppler maps implementing the
maximum likelihood approach. As test data, we used observations of the dwarf
nova V2051~Ophiuchi. The system was observed in quiescence at least 16~d before
the onset of the next outburst. Using Doppler maps obtained for ten emission lines
covering three orbital cycles, we detected spiral structures in the accretion disc of
V2051~Oph. However, these structures could be biased as our data sampled the orbital
period of the binary at only eight different orbital phases. 
Our Doppler maps show evolution from a one-arm wave structure in H$\alpha$ to two-armed waves in the other lines.
The location of the two-arm structures agrees with simulations showing tidally driven spiral waves in the accretion disc.
During consecutive cycles, the qualitative characteristics of the detected structures remained similar but the central absorption increased.
For the first time, using the Doppler tomography method, we obtained temperature maps of the accretion disc.  
However, taking into account all the assumptions involved when using our method to retrieve them, 
the result should be treated with caution.
Our maps present a relatively flat distribution of the temperature over the disc, showing no temperature increase at the location of the spiral arms.
Using `ring masking', we have revealed an ionized region located close to 
the expected location of stream--disc interactions.
We found the average temperature of the accretion disc to be 5600 K, which is below the critical limit  deduced from the disc instability model.
\end{abstract}




\section{Introduction}
Accretion discs are important in astronomy because they drive the activity of various astrophysical objects and environments, such as cataclysmic variable stars, X-ray binaries, supernovae, active
galactic nuclei and others \citep{frank2002}.
\citet{shakura1973} introduced a phenomenological prescription of the standard model of accretion discs. Since then, huge progress has been made in observational and theoretical studies of accretion \citep[e.g.][and references therein]{blaes2011}. 
However,
some of the most fundamental questions regarding accretion discs and their role in the evolution of systems that host such discs have not been answered to date.
Because of their minuscule angular dimensions, accretion discs cannot be spatially resolved, 
prompting intensive efforts to find methods that can overcome this difficulty.  
To probe structures on angular scales of microarcsec, researchers have implemented various tomographic methods, such as eclipse mapping \citep[e.g.][]{horne1985} and Doppler tomography \citep[e.g.][]{marsh1988}. 
Several interesting results have been obtained with these techniques.  
The most important seems to be the discovery of two-armed spiral shocks in the accretion disc of IP Peg \citep{steeghs1997}. However, there are also other notable results such as the detection
of a bright spot at the stream--disc impact region in cataclysmic variable stars  \citep[e.g.][]{bloemen2013,savoury2012} or probing the emission from the secondary star in X-ray binaries \cite[e.g.][]{zurita2016}. An introduction to different tomographic
methods and their applications can be found in \cite{boffin2001a}.
Here, we present the capabilities of our code for Doppler tomography, which has been tested using the spectra of
the well-known dwarf nova V2051~Ophiuchi. We also show how one can try to estimate the temperature of the structures revealed by
Doppler tomography.    
 
Dwarf novae are a subclass of cataclysmic variables, which, because of their prevalence, short orbital periods and repeated outbursts \citep[e.g.][]{warner2003}, are a perfect laboratory to study the accretion process. 
Such systems contain a white dwarf accreting material from a cooler companion star. Because the matter from the companion has a significant angular momentum, it cannot flow directly on to the white dwarf, but instead forms an accretion disc. Accretion discs are highly variable in brightness, featuring repetitive outbursts and superoutbursts. Extensive photometric observations of dwarf novae have contributed a lot of information to help us to understand these phenomena \citep[e.g.][]{rutkowski2009, olech2011,otulakowska2013}. 

Observations of V2051~Oph were first mentioned by \citet{Sanduleak1972}. 
Its orbital period was measured by several authors \citep[e.g.][]{echevarria1993,brunt1983,baptista2003}, who found $P_{\rm orb}\approx 89.9$ min.
V2051~Oph shows recurrent outbursts with a 2--3 mag amplitude. During superoutbursts it reaches $V\sim12$ and during 
quiescence its brightness drops to $V\sim15.5$ \citep{kato2001,baptista2007}. 
Using data from the American Association of Variable Star Observers (AAVSO), one can also estimate the period between consecutive
superoutbursts and normal outburst to be $\sim300$ days and $\sim50$ days, respectively.  
 V2051~Oph belongs to short-period CVs with deep eclipses \citep[e.g][]{savoury2011}. Eclipses in V2051 (of an amplitude $B\sim2.5$) are caused by its high orbital inclination ($i\sim83.3$ deg.)
Therefore, it was possible to study this object taking advantage of the eclipse mapping technique
 \citep[][]{vrielmann2002,baptista2004,baptista2007}.  Moreover, spectroscopic studies revealed double-peaked line profiles in the emission spectrum of V2051~Oph. As a result, it was possible to use Doppler tomography to study this system. \citet{papadaki2008} presented their Doppler maps revealing ring-shaped  emission of the accretion disc with two main features: a bright spot and a possible region of a superhump source. Recently, \citet{longa-pena2015} presented their maps pointing to similar emission regions in the disc, though finding different properties for them. Moreover, they aptly detected in their spectra a strong Ca${\rm \ II}$ emission triplet, which was used as an indicator of emission from the mass donor component. 
Because of this, they found a mass ratio $q=0.18\pm0.05$ and a semi-amplitude of the radial velocity (RV) $K_1=97\pm10 {\rm\ km}{{\rm\, s}^{-1}}$, which is in agreement with previous estimations. 
\section{Observations and reduction}\label{observations}
\begin{figure*}
\includegraphics[width=0.35\textwidth,angle=-90]{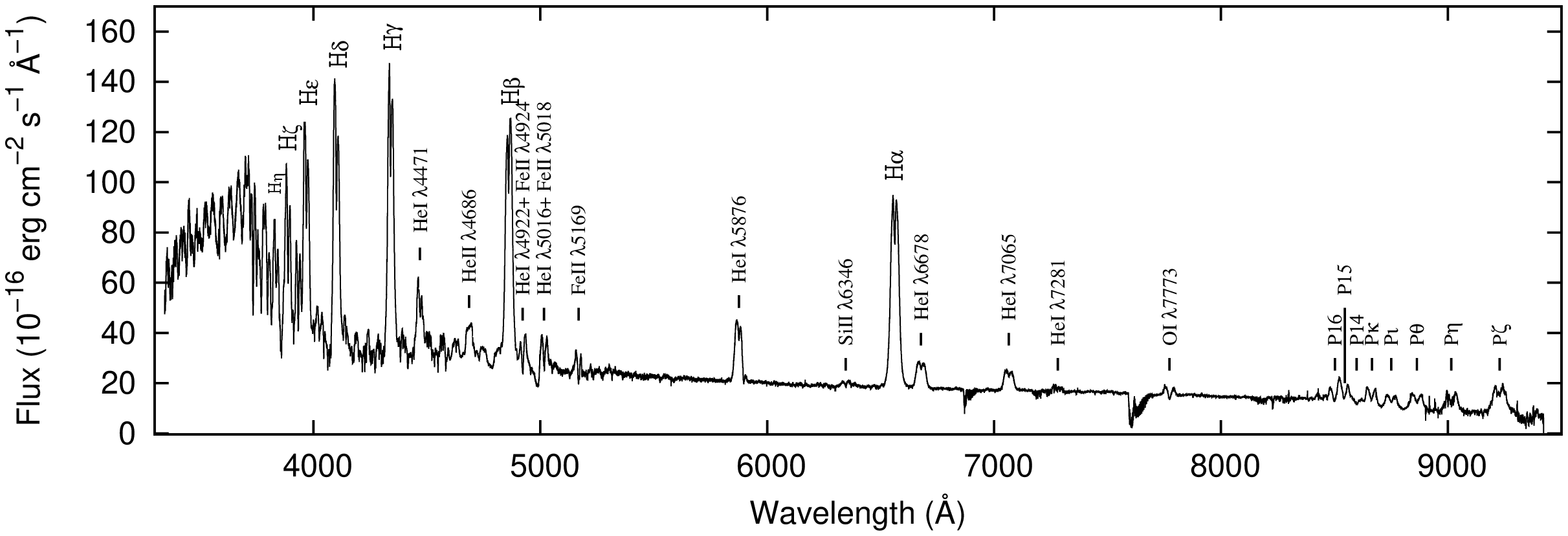}
\caption{Average spectrum of V2051~Oph composed of 21 spectra. The most prominent lines are marked on the graph.} \label{fig:averaged}
\end{figure*}
We made observations of V2051~Oph covering nearly three orbital periods on 2005 August 21. We used the MIKE echelle spectrograph \citep{bernstein2003} mounted on the 6.5-m Magellan II (Clay) telescope located at the Las Campanas observatory, La Serena, Chile.  
 We obtained 21 spectra of V2051~Oph  with the blue and red arm simultaneously. Each exposure was obtained with a 600 sec integration time.
The MIKE spectrograph delivers full wavelength coverage: 3350-5000 \AA \ in blue and 4900-9500 \AA \ in red arms. We used a 0.75 arcsec slit, which gave us a resolving power R$\sim$49000 (6.1 km s$^{-1}$) in the blue side of spectra and R$\sim$38000 (7.8 km s$^{-1}$) in the red side. Both the red and the blue arm detectors have 2048x4096 15$\mu$ pixels. When using 2x2 binning, this results in a dispersion $\sim$ 0.1\AA / pix. 
Cosmic rays were removed from the images using the {\it dcr} software described by \cite{pych2004}. 
The spectra were processed with a pipeline developed by Dan Kelson following the formalism of 
\cite{kelson2003}. The star HR 5987 was observed on the same night and used as a spectroscopic standard during reduction. We selected this star because of its early spectral type
B2.5Vn. Because the spectrum of V2051 Oph has relatively broad features, blaze functions for the echelle apertures were estimated
based on the continuum fitting to the spectrum of our standard. A useful
wavelength range for each aperture was selected using the {\it contlim}\footnote{http://users.camk.edu.pl/pych/ECHELLE/index.html} software. We rejected the apertures which did not overlap with the neighbors in the wavelength coverage.
We multiplied our reduced spectra by the calibrated spectrum of HR 5987 taken from the ESO Science Archive\footnote{http://archive.eso.org/cms.html} in order to calibrate our spectra in erg. 

During the observations, the sky was covered with smooth cirrus, but we found that the weather, although not photometric, was
sufficiently stable. However, in order to exclude that small variations of the continuum level are color dependent (which would result in a distortion of line profiles and their relative intensities), we performed the following procedure. 
First, we chose a few of the best spectra (in terms of the most stable atmospheric conditions and the
highest signal) observed directly after the HR 5987 calibrator in order to construct the mean reference spectrum. Each previously
calibrated spectrum was divided by this reference spectrum. As a result, we obtained auxiliary spectra free of evident emission or absorption
features, reflecting the ratio between consecutive spectra and the reference spectrum. The obtained auxiliary spectra were in
general flat, meaning that the temporal changes of absorption had a `gray' character (i.e. were wavelength-independent). Therefore,
such changes were likely to be caused by slit losses and clouds.We
fitted second-order polynomials to all the auxiliary spectra. Next,
all the calibrated spectra were divided by the corresponding fits.
Finally, using this procedure, all the spectra have been normalized
assuming that the extra atmospheric continuum should have
the same level all the time. Note that the above procedure tries to
compensate for the effects coming from variations due to clouds
and unstable seeing. However, it does not take into account source
variations during eclipses. Although, in general, our data have been
obtained outside an eclipse, the spectrum for phase 0.98 of the second
observed orbital cycle is suspected to be biased by an eclipse
(it was excluded from the Doppler tomography). Although we have
done as much as possible to properly calibrate our spectra, taking
their quality into account, the resultant flux shown in Fig.~\ref{fig:averaged} could be
erroneous by a few tens of per cent, and therefore should be treated
with caution.

 According to AAVSO, our observations (JD 245~3603) caught V2051~Oph about 150 d after the prior outburst and $\sim16$~d before
the next outburst. During this time, the system had $\sim$15.0--15.5 mag.
\citet{longa-pena2015} carried out observations (JD 245~5386) $\sim$4~d after the outburst maximum when the system had $\sim$13.5 mag. \citet{papadaki2008} performed observations (JD 245~1394) around 8~d after the outburst maximum. During their observations,
V2051 had 14.5-15.0 mag. Taking the above facts into
account and examining the light curve, it is obvious that our observations
were the closest to quiescence among the observational
runs mentioned above. However, careful investigation of the light
curve reveals that it is not the lowest photometric state in which the
system can be found. Certainly, V2051 was neither in outburst nor
in rise to outburst during our observations.
\section{Spectrum} \label{spectrum}
\begin{figure*}
\includegraphics[width=0.7\textwidth,angle=-90]{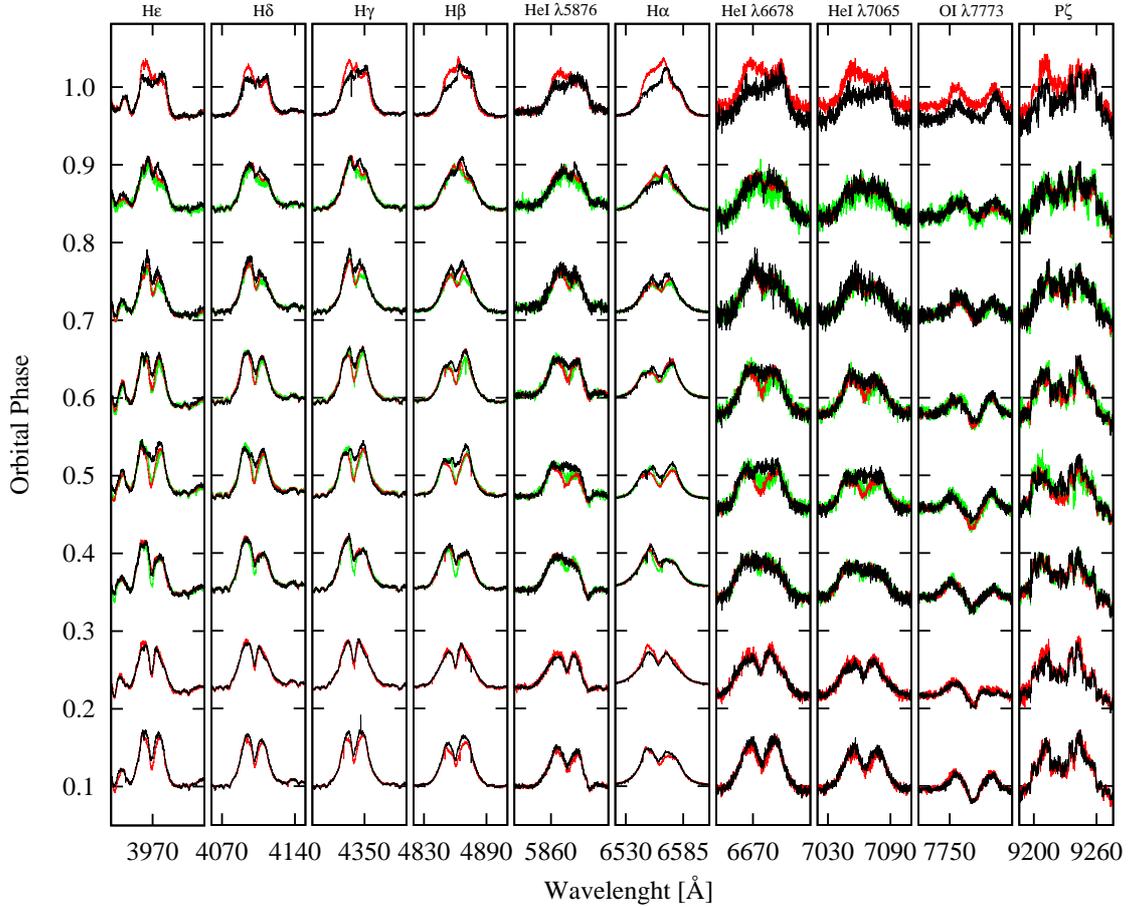}
\caption{Profiles of ten prominent emission lines present in the spectrum of V2051~Oph. Profiles were ordered according to the approximate orbital phases
(see text). Black, red and green line colours denote the first, second and the third orbital cycle, respectively. Relative line intensities are presented.}  \label{prof_together}
\end{figure*}

Figure~\ref{fig:averaged} presents the average out-of-eclipse spectrum of V2051~Oph where we have marked emission lines.   
The most prominent lines belong to the Balmer series, however there are also noticeable lines of 
HeI, HeII and FeII. 
This characteristic is consistent with the \citet{papadaki2008} observations. However, our spectra have a wider spectral range (beyond 7500\AA).  On the red side of the spectrum, one can see weaker but still clearly visible emissions from Paschen series, Si~II$\lambda$6346 and O~I$\lambda$7773. We were not able to find emission from
CaII as reported by \citet{longa-pena2015}. Our spectra are of higher resolution and better signal-to-noise (S/N)
than both results mentioned above. Hence, the absence of the CaII emission must result from the fact that our observations caught the system in a lower state in comparison to \citet{longa-pena2015}.
Clearly, a much weaker flux from the inner accretion disc was not able to excite  
 CaII emission from the secondary. All Balmer and HeI lines have stronger blue side peaks except of H$\beta$ and HeI$\lambda$4922, where red sides are stronger. 
A close view of the proximities of ten most prominent line profiles is presented in Fig.~\ref{prof_together}.
The spectra were phase-ordered according to the photometric ephemeris obtained by \citet{baptista2003} and corrected for the spectroscopic phase shift. 
The profiles have been cleaned out of narrow telluric emission and absorption lines.
Figure~\ref{prof_together}. shows that the levels of the blue and red peaks are correlated with the orbital phase.  

The fact that the ratio of the orbital period to the time duration between neighbouring exposures is roughly an integer number resulted
in overlapping of the observed orbital phases in consecutive orbital cycles. Therefore, although we covered nearly three orbital
periods with our observations, the system was observed only around eight different orbital phases. However, this circumstance has one
advantage as it allows for a direct comparison of different orbital cycles. Small differences in the orbital phases were omitted in Fig.~\ref{prof_together} in order to facilitate a comparison of the line profiles. 
Inspection of Fig.~\ref{prof_together} reveals changes of the line profiles from phase to phase,which
are induced by the orbital motion of the system. However, we can see that for some emission lines, even for close phases of different
cycles, the profiles are not identical. This shows that the accretion disc is not entirely in a steady state and/or the outer regions of the accretion disc are populated by cooler matter with significant optical depth incidentally obscuring light-emitting regions. Fluctuating
intensity of the Balmer lines for the same phases of different periods has been also reported by \cite{papadaki2008}. They found the same explanation for this phenomenon.  At the phase 0.98 
(which is close to the spectroscopic eclipse), the shapes of the line profiles are significantly different in comparison to profiles observed at other phases. In particular, central absorptions are practically invisible. In some cases, such as for H$\alpha$ and H$\beta$, those absorptions convert into emissions but it should still be remembered that the actual line flux is much lower during eclipse phases than outside them. Because at this phase the inner disc is obscured by the secondary, the above emission may originate from its surroundings.
\section{Radial velocities} \label{radial}
\begin{table*}
\centering
\caption{Orbital parameters derived from fitting equation (2) to RV measurements. Uncertainties  are calculated as a standard errors in non-linear least-squares fitting.  See text for details.
}
\begin{tabular}{@{}l|rrrrrr@{}}
Cross-correlation   & $\gamma$ & err $\gamma$ & $K$  & err $K$ & $\phi_0$     & err $\phi_0$ \\ \hline
H$_{(\alpha,\beta,\gamma,\delta,\epsilon)}$       & --  & --        & 63.5 & 4.4  & 0.0922 & 0.0119  \\
HeI$_{\lambda 6678,\lambda 7065}$                   & -- & --      & 85.0 & 4.6 & 0.1103  & 0.0081 \\
FeII$_{\lambda 5169}$                                & --   & --        & 89.9  & 5.5  & 0.0822   & 0.0995    \\
\multicolumn{7}{c}{} \\
Mirroring  & \multicolumn{6}{c}{} \\ \hline
H$_\alpha$   & -55.7 & 3.8        & 79.1 & 5.1 & 0.1020  & 0.0110   \\
H$_\beta$    &\ 21.3  & 5.3        & 83.7  & 7.2  & 0.0815 & 0.0147   \\
H$_\gamma$   &\ 31.6  & 5.3        & 53.0 & 7.2 & 0.0768 & 0.0231   \\
H$_\delta$   & -39.1 & 3.5        & 65.5 & 4.7 & 0.0894 & 0.0122   \\
H$_\epsilon$ & -137.3 & 4.3        & 51.9 & 5.9 & 0.1019  & 0.0193  \\
\label{tab:rv}
\end{tabular}
\end{table*}
\begin{figure}
\includegraphics[width=0.49\textwidth]{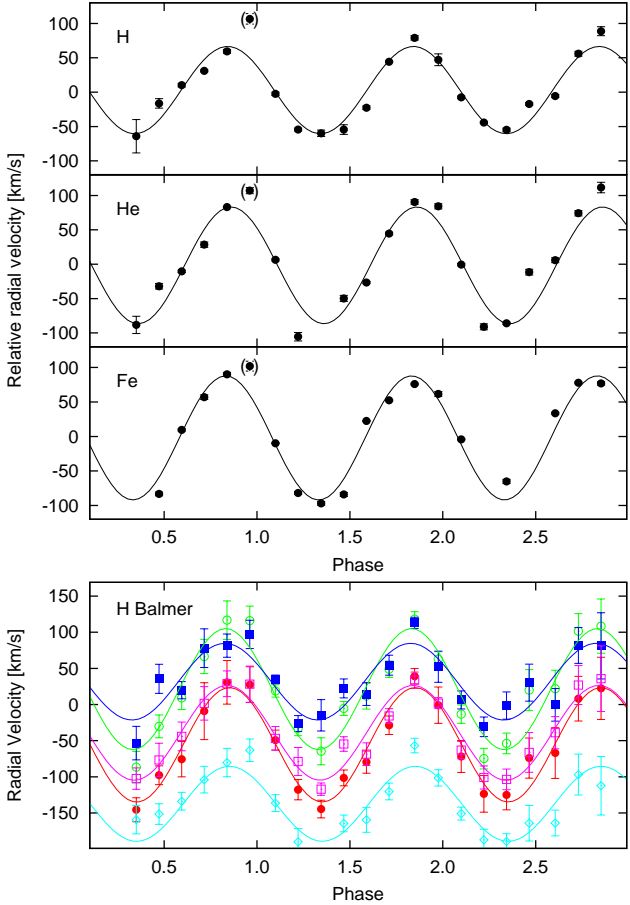}
\caption{RV curves with corresponding sinusoidal fits shown for almost three observed periods. The top three panels show the results from the cross-correlation approach. Solid lines represent best fits with parameters
shown in Table~\ref{tab:rv}. Points in parentheses were excluded from the fits. 
The bottom panel shows RV measurements given by the method of mirroring.
The colors represent H$\alpha$ (red), H$\beta$ (green), H$\gamma$ (blue), H$\delta$ (purple), H$\epsilon$ (cyan).
} \label{fig:radial}
\end{figure}
In our subsequent analysis involving RV measurements we have used the following orbital ephemeris 
 derived by \citet{baptista2003}.
 \begin{equation}
T_{0}[HJD]=2\ 443\ 245.9769 + 0.062427860E
\end{equation}
Correct measurements of the RVs are extremely important for an accurate determination of system parameters. Previous RV studies
provided significantly different values of the system velocity $\gamma$ for different emission lines \citep[e.g.][]{watts1986,longa-pena2015}. 
We have measured the RV curve of our system using three approaches: cross-correlation, mirroring and fitting the central dip in the line profiles. All measurements were then fitted by the following formula:
\begin{equation}
v_{r}(\phi)=\gamma-K \sin(2\pi(\phi-\phi_{0})) \label{eq:rv}
\end{equation}
where $\gamma$ is the systemic velocity, $K$ is the semi-amplitude of the RV curve, 
$\phi$ is the orbital phase and $\phi_0$ is the phase shift. 

The first method we used to determine RVs involves the cross-correlation technique. We correlated five H lines ($\alpha, \beta, \gamma, \delta, \epsilon$), two HeI lines ($\lambda6678$, $\lambda7065$), and one FeII line ($\lambda5169$) with appropriate templates obtained by averaging the line profiles of all individual spectra. After obtaining RVs, we rescaled each spectrum to put it at arbitrary zero velocity, then we prepared new, improved templates and repeated the cross-correlation measurements. This method brings us information about the RV amplitude and phase shift only.

The second approach involves mirroring of the line profiles and is especially useful when one works with asymmetric and complex
line profiles such as ours. Using this method, we focused only on the wings of the profiles. It is commonly accepted that they are the best
indicators of the RV of the primary component because the outer parts of spectral line profiles generally involve central regions of the
accretion disc directly surrounding the primary. Unfortunately, this method is very sensitive to even weak blends present in the wings
of the line profiles. This implies that the measured system velocities $\gamma$ have different values for different emission lines (as shown by several authors). The RVs obtained using the above two methods and the sinusoidal fits based on the parameters from Table~\ref{tab:rv} are presented in Fig.~\ref{fig:radial}.

The third method relies on fitting a polynomial to the dip located between side peaks of the line profile and finding the position of
the minimum. \cite{watts1986} reported that the position of the central minimum in their data did not show any periodic variation
consistent with the orbital motion. However, in the case of our data, the situation looks much more promising. Although both the amplitude $K_1$ and the phase shift $\phi_0$ are determined with low accuracy, the third approach gives the most consistent estimation of the system velocity $\gamma$ amongst the considered methods.
What is more, all the obtained values  of $\gamma$ (from H$\alpha$ to H$\epsilon$: $-20.57, -37.31, -23.7, -0.75, -50.5$) suggest that the system approaches the observer.
The average value of the systemic velocity $\gamma=-26.6\pm18.7$ km~s$^{-1}$. We adopted this number in further analysis. 
  
The hydrogen lines give the most reliable estimation of $K_1$ for H$\alpha$ and H$\beta$. 
Because of the lower S/N ratio and unknown influence from blends, higher-order Balmer lines give a less certain estimation. HeI($\lambda6678$), HeI($\lambda7065$) and FeII($\lambda$5169) lines are located in the smooth part of the continuum, and therefore they also give good estimations of $K_1$.
Hence, to obtain average values of $K_1$ we took into account the HeI and FeII estimations from the cross-correlation and H$\alpha$ and H$\beta$ from the mirroring method. In this way, we obtained the semi-amplitude for the primary component to be $K_1=84.4$~km~s$^{-1}$ with the sample standard deviation $4.4$~km~s$^{-1}$. 
This value is in agreement with previous estimations by \cite{baptista1998}, \cite{papadaki2008} and \cite{longa-pena2015}. As the final value of $\phi_0$ we take the average of the cross-correlation and mirroring methods i.e. $\phi_0=0.092$ with the sample standard deviation $0.012$. This was used to correct the photometric ephemeris and phases for Doppler tomography.
\section{Doppler tomography} \label{dopplertomography}
The main source of the broad emission lines of V2051 Oph and other cataclysmic variables is gas moving around the white dwarf
with nearly Keplerian velocities. This gas creates an accretion disc. The other much less prominent sources may be connected with the
irradiated secondary star, the stream of matter coming from the secondary and/or from the corona located above the accretion disc. As
previously shown, the emission lines in V2051 Oph have a complex highly time-dependent profile structure. Line profiles observed at
each orbital phase provide projection of these structures along the line of sight. This observational effect has become a foundation of
the Doppler tomography technique developed by \citet{marsh1988}. There are several widely used codes for calculating DMs (such as 'MOLLY' created by T.~Marsh)  implementing the Fourier-filtered back-projection (BP) approach. Others, such as the one created by \citet{spruit1998}, implement the maximum entropy method (MEM). 
 
We have developed a new code that converts time series of the line profiles into DMs.
Our code implements the maximum likelihood (ML) approach \citep{richardson1972,lucy1974}. For the Poissonian signal statistics, which is a typical situation for astronomical data, this algorithm maximizes the likelihood function given by the expression:
\begin{equation}
\sum_{i=1}^{I} \sum_{j=1}^{J} S_{ij} \ln \left( \sum_{k=1}^{K} \sum_{l=1}^{L} P_{ijkl} D_{kl} \right)
\end{equation}
Here, $S_{ij}$ is the observed spectrum signal for the velocity bin $i$ and the orbital phase $j$, $P$ is the tensor describing projection from the DM to the observed line profiles ($P_{ijkl}$ shows how the brightness in a given pixel of the map influences the spectral line profile at a given velocity bin and orbital phase of the system), $D_{kl}$ is the brightness signal for the pixel $kl$ of the DM. As we can see, to find the DM of a size $K \cdot L$ pixels we need line profiles having $I$ velocity bins for $J$ orbital phases of the system.
The above maximization condition leads to the well-known iterative scheme \citep{richardson1972,lucy1974} which enables to obtain the DM in a suitable but relatively time-consuming way. Years ago, we used a similar approach with success to infer the directional distribution for the dust outflow from a cometary nucleus using a series of images of the cometary dust coma \citep{waniak1998}.

The serious issue with our data is that we have sampled the orbital period of the binary at only eight different orbital phases,
which could seriously bias the obtained DMs. However, the integration time for all the spectra was close to the time delays between
subsequent exposures (typically slightly shorter by an overhead and readout time). As a result, even a properly reconstructed DM will be
azimuthally smeared by roughly 40 deg. This circumstance, however, enables us to introduce a number of supplementing line profiles
at phase points between the observed phases.We have used a simple linear interpolation in the phase space between the existing profiles.
This way of interpolation guarantees that no false structure in the DM can be created. Although our approach seems to be ad hoc, it
works quite well. Indeed, if the DM contains circularly symmetric structures such as a homogeneous accretion disc, our algorithm ensures
a perfect reconstruction of the DM even from one line profile, which can be duplicated as many times as necessary. However, if
we have strictly radial features our algorithm reconstructs them by introducing additional azimuthal smearing but below the smearing
produced by the integration time. In any case, the suppressing of reconstruction artefacts provided by our `phase multiplication by
interpolation' approach is evident. From trial-and-error test computations, we have concluded that a total number of 33 line profiles
(including the eight original ones) for different phase points ensures perfectly good quality and reliability of the DMs as well as efficient
computations at the same time. 

With the aim to test our algorithm, we have performed a number of reconstructions of DMs from artificially prepared line profiles as
well as from observed line profiles using the classical filtered BP method and our ML algorithm. Our typical test input images consisted
of an azimuthally symmetrical disc of different radial profiles and a couple of disc structures such as hot spot, arcs and
spirals. Such images underwent angular smearing introduced by our integration time. Then, we computed the spectral profiles and
degraded them by noise at a level typical for the S/N ratio in our observed profiles. After that, the DMs were obtained and compared
with the azimuthally smeared input profiles. For the BP scheme, we have used a filter passing as much high spatial frequency signal
as possible from the point of view of the reconstruction stability. Likewise, using the ML approach we have applied as many iterations
as possible yet ensuring a stable reconstruction. To increase the smoothness of the presented DMs, we convolved them with a
circularly symmetric bell-shaped function of an appropriate, experimentally obtained FWHM. We have found that the function width
of a few velocity pixels is quite enough to produce smooth images, not influencing so much the features existing in the DMs.

\begin{figure*}
\includegraphics[width=1\textwidth]{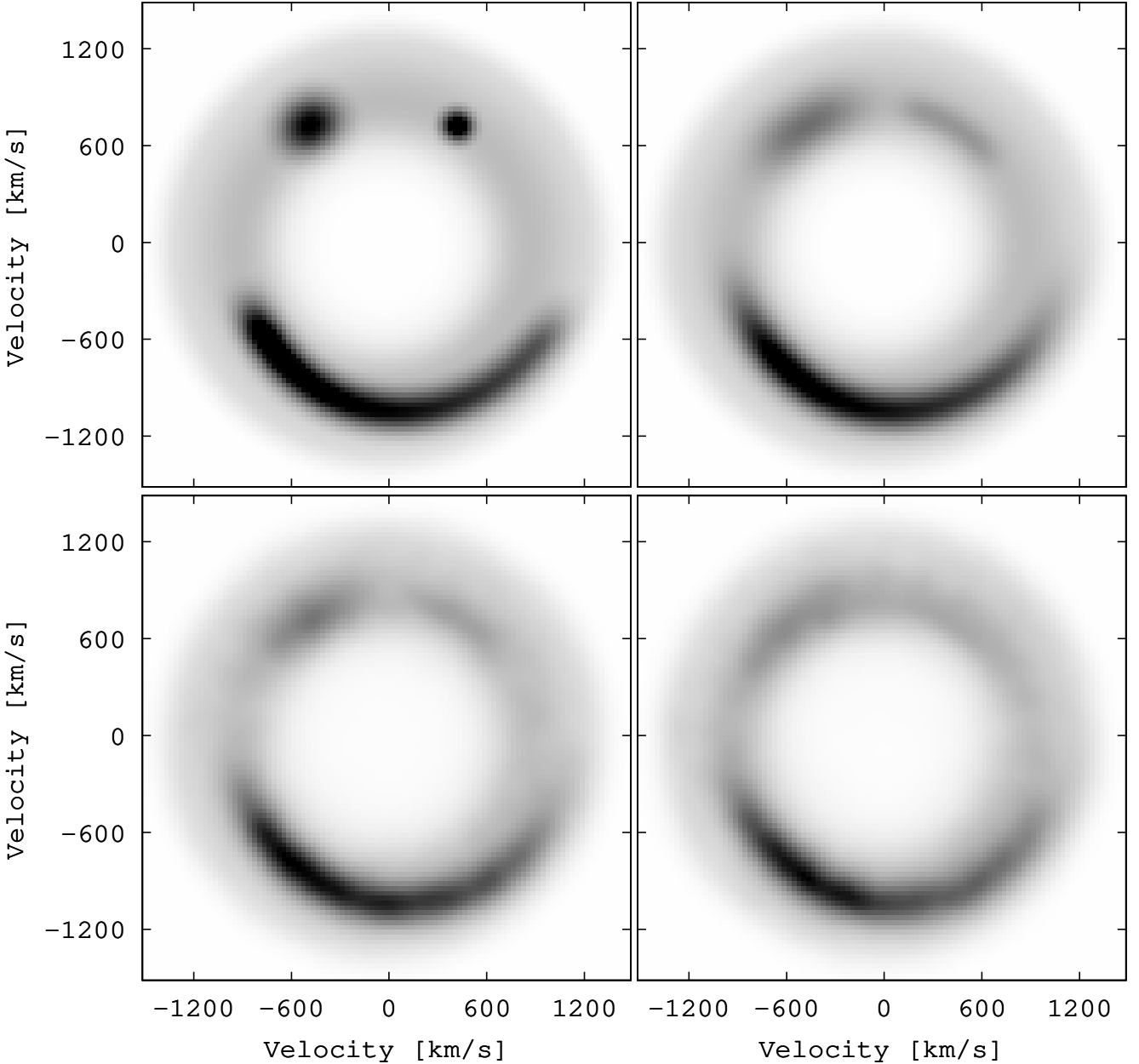} 
\caption{Result of one of our test computations presenting possibilities of our ML approach to the Doppler tomography combined with our method of phase
multiplication by interpolation (see text for details). Top-left panel: original input map of the Keplerian accretion disc with the temperature profile T$\sim$R$^{-3/4}$, harbouring two hot spots of different sizes and a part of a spiral. Top-right: input map azimuthally smeared according to our integration time. Bottom-left: DM
reconstructed from the 33 spectral profiles registered at different phases. Bottom-right: DM computed from the eight profiles obtained at exactly the same
phases as our data for V2051~Oph. The 25 missing profiles have been computed using our interpolation method.} \label{interpolation}
\end{figure*}

Here we present one example of our test computations showing how our ML approach can reconstruct the DM from eight orbital phases (Fig.~\ref{interpolation}).
The input map of the disc underwent the angular smearing introduced by the integration time. Then it was used to produce a number of spectral profiles at different phases. The set of such profiles, degraded by noise, was an input for retrieving the DMs. As can be expected, the smallest spot is completely degraded mainly by the angular smearing. The restricted number of the input orbital phases biases its profile much less effectively. The spiral structure, although also distorted, looks quite similar to the original input structure.

\begin{figure*}
\includegraphics[width=1\textwidth]{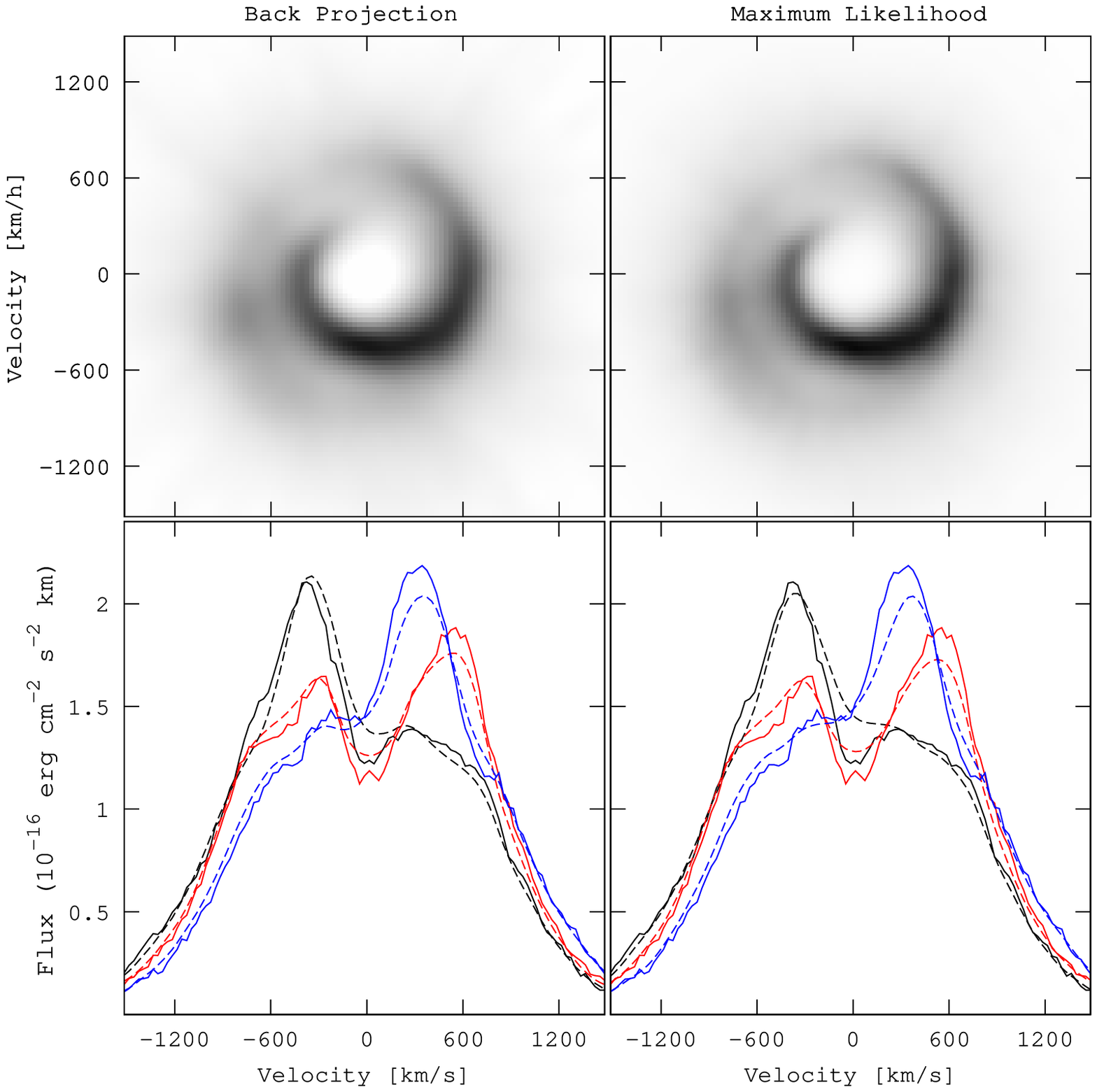} 
\caption{Comparison of the results given by the BP and ML methods. Top panels: DMs for H$\alpha$ emission of V2051 Oph. Bottom panels: observed line profiles (solid curves) and corresponding output profiles computed from the reconstructed DMs (dashed curves). We present only three exemplary phases in order to improve clarity of the graph.} \label{comparison}
\end{figure*}

In order to compare how accurate our BP and ML methods are, Fig.~\ref{comparison} presents DMs for H$\alpha$ emission of V2051 Oph for
our first orbital period as well as the observed and computed line profiles. We transferred the original line profiles to velocity bins each having 30~km~s$^{-1}$, the same as the size of the square pixel in the DMs. (The motivation of this width of velocity bin is presented in Sec.~\ref{dopplermaps}.)
We can see a distinct spiral wave feature in the DMs given by both methods. What is more, differences between the results are small. The BP method gives a little more patched structure, whereas the ML approach tends to transfer the signal from the side peaks of the line profile to the central dip. This is a quite obvious result of the intrinsic data treatment by the two methods. The ML method gives smoother DM maps and theoretically recovered line profiles compared to the BP algorithm.
\section[]{Doppler maps} \label{dopplermaps}
\begin{figure*}
\includegraphics[width=1\textwidth]{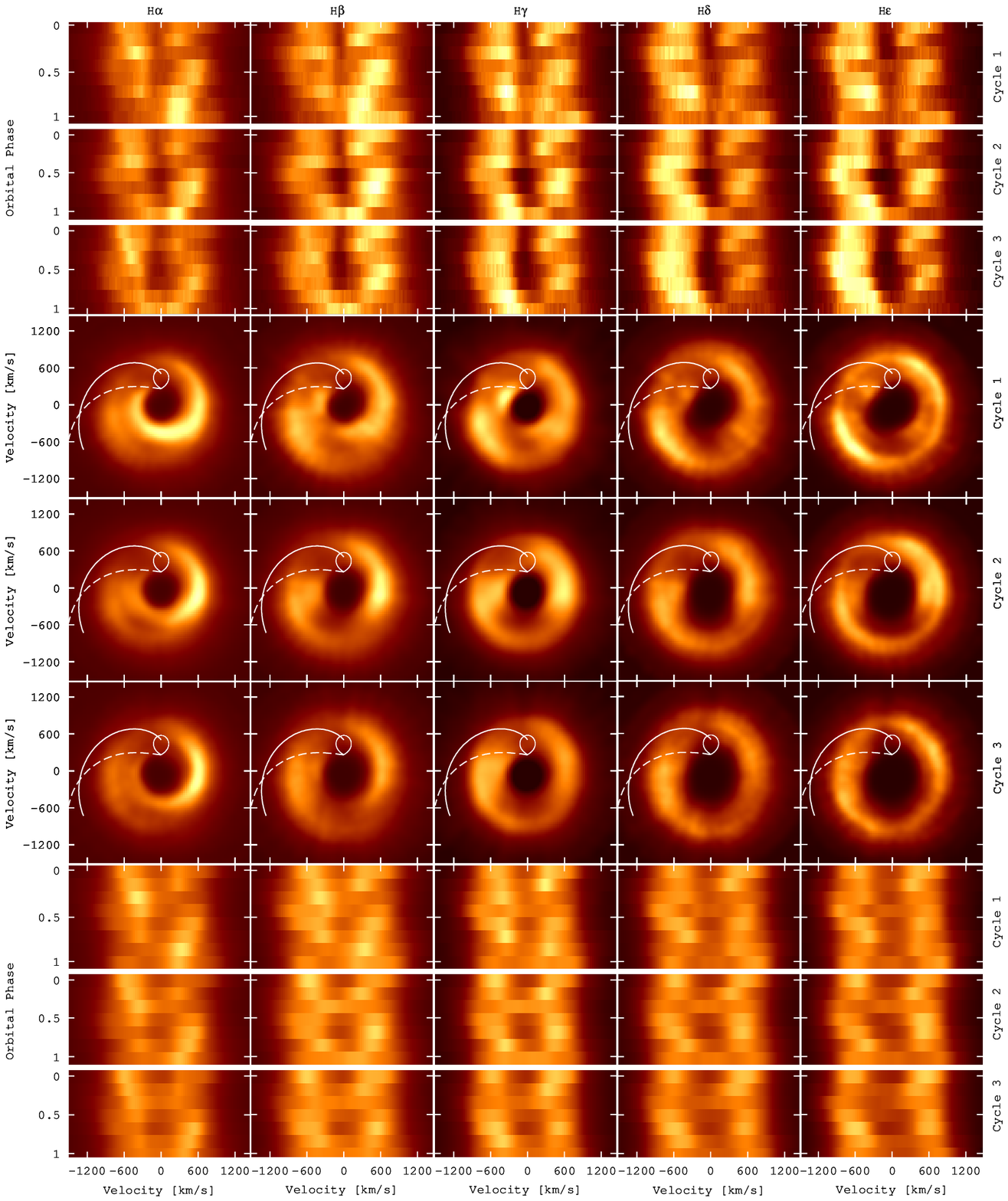} 
\caption{DMs and trailed spectra for five Balmer lines obtained for three orbital cycles (see text). The first three rows present input observed trail spectra, the
three middle rows show DMs and three bottom rows show the trailed spectra recovered from the DMs. Time goes from top to bottom. The brightness scale is
the same for each particular column, but may change along a row in order to improve visualization.}
\label{doppler1}
\end{figure*}

\begin{figure*}
\includegraphics[width=1\textwidth]{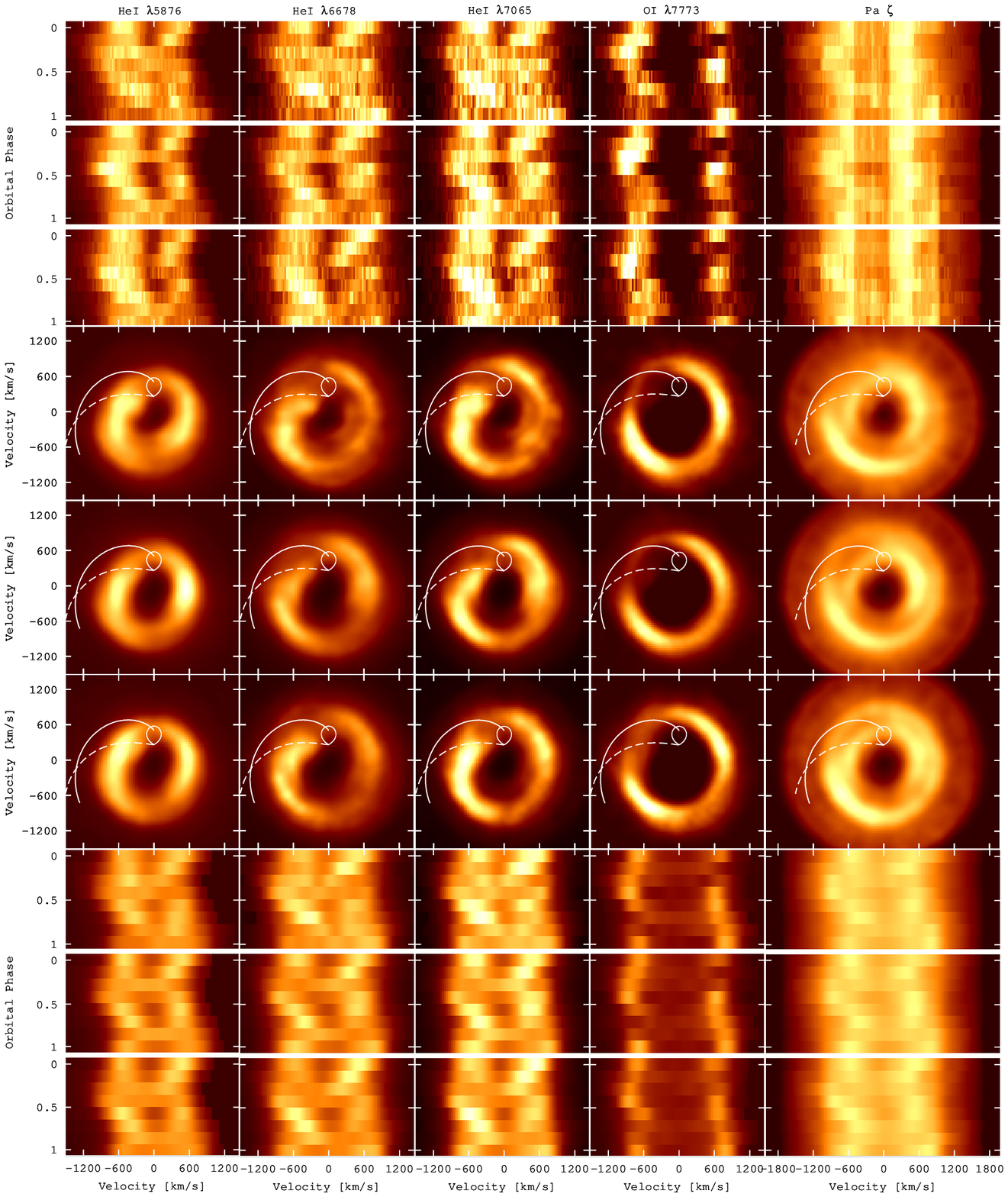} 
\caption{Doppler tomography maps and trailed spectra obtained for emission lines of HeI, OI and Pa$\zeta$. The figure is organized in a similar way as Fig.~\ref{doppler1}.}
\label{doppler2}
\end{figure*}

One of the principles of the Doppler tomography is that the flux of each element in the DM remains constant in time or at least during
the one orbital period analysed. This means that there is a constant total flux in all the observed line profiles for this period. Therefore, in order to fulfil this condition, each line profile has been normalized to the average total flux over all phases of the cycle. Then, the profiles were transformed from wavelengths to velocities and re-binned to the velocity channels having a width of $30~{\rm km}~s^{-1}$. These profiles are presented as trailed spectra in Figs~\ref{doppler1} and~\ref{doppler2}.  As was mentioned 
in Section~\ref{observations} our spectra have theoretical resolution of the order of 10~km~s$^{-1}$. Unfortunately, using such a high resolution of the line profiles does not enable us to reconstruct the DM with similar resolution, as our original data involve only eight different phases per one orbital period. Even our method of phase multiplication by interpolation does not overcome this difficulty. Thus, using velocity bins of the width of 30~km~s$^{-1}$ is more adequate from the point of view of the data structure, and this produces line profiles with an increased S/N ratio. 

We obtained Doppler maps for ten emission lines presented in Fig.~\ref{prof_together}. The other emission lines visible in Fig~\ref{fig:averaged}, although sometimes quite strong, were omitted because of the biasing by blends and/or the difficulty in proper determination of the continuum level.
Figures~\ref{doppler1} and~\ref{doppler2} show Doppler Maps obtained using our ML implementation of Doppler tomography together with corresponding trailed spectra for the Balmer series (Fig.~\ref{doppler1}) and five lines of HeI, OI and Pa$\zeta$ (Fig.~\ref{doppler2}) respectively.
 Our observations cover over 2.5 orbital periods. Therefore, we have the opportunity to show how the DMs vary from cycle
to cycle. To do this, we divided our data into three consecutive blocks, each encompassing one orbital cycle and giving one DM. Of
course, we did not avoid the partial overlapping of the neighboring orbital periods. For the Balmer lines, the first block includes first eight exposures 1-8, which covers one whole orbital period, while second and third blocks include seven exposures 7-13 and 15-21.
Although all our spectra have been obtained outside the eclipse,
the exposure 14 close to phase 0.98 having an entirely different character, was excluded from the Doppler tomography analysis. Its line profile is shown by the red line in Fig.~\ref{prof_together} and was discussed in Sec.~\ref{spectrum}.
TThus, for the second and third orbital periods,DMs have been
reconstructed from seven different original phases (interpolated, as previously, for additional 26 phases). Therefore, we performed the test in which we reconstructed the DM for the first cycle excluding one phase nearest to the phase 0.98 and compared it with the DM
obtained from the eight original phases. On average, discrepancies between both results were of the order of a couple of per cent, not
exceeding 10 per cent.

For the five lines of HeI, OI and Pa$\zeta$ we also organized the input line profiles in three blocks each having eight phases. The first
block includes exposures 1--8, the second comprises exposures 7--14 and the third contains 13--20. After careful checking, we excluded exposure 21 as its S/N ratio was the worst and the overall structure for all five line profiles strongly deviated from the rest (the reason remains unknown).
The overall layout of Fig~\ref{doppler1} and Fig~\ref{doppler2} is the same. The first three rows show the observed trail spectra. The data blocks and profiles are ordered in time from top to bottom. The three middle
rows present DMs and the bottom three rows show the trail spectra reconstructed from the DMs. They are organized in the same way
with respect to time (from top to bottom).All DMs are centred on the mass centre of the system.
In order to facilitate further analysis, the DMs were supplemented with a Roche lobe of the secondary star. The velocity curve of the Keplerian disc along the gas stream and the ballistic trajectory of this stream have been shown as the upper solid and lower dashed lines, respectively. The Roche lobe of the secondary and both trajectories have been plotted according to the system parameters derived by \citet{baptista1998}.
The trajectories were calculated in the corotating frame of the velocity space after taking rotation of the whole system into account: $\widetilde{V}_X=V_X-\omega Y$, $\widetilde{V}_Y=V_Y+\omega X$, where $X,Y$ are positions in the corotating frame, $\omega$ the angular velocity.
The display scale of signal intensity is the same for each particular column, but it can change from one column to another to be properly adjusted in order to improve visibility of the features.

Small intensity changes for the same phases belonging to the consecutive cycles can be seen in Fig.~\ref{prof_together}. One can expect that such brightness variations are signs of superhumps or flares in the accretion disc. Because these variations are very small, they violate
the assumption about the stability of the disc only marginally. So, neither the trailed spectra nor the DMs are affected significantly by
them. 

All the trailed spectra show a clear double-peaked structure of the emission line. The blue peak is the eclipse prior to the red
peak. As was previously mentioned by \citet{papadaki2008}, this indicates prograde rotation of the disc. There is no clear evidence
of the S-wave and the overall structure of the trail spectra is quite complex, indicating the existence of multiple sources of emission.
We can also follow how the trail spectra change from cycle to cycle. For all the lines, the central absorption becomes stronger as time
goes on. Evolution of the central absorption in the trailed spectra of the Balmer series (see Fig.~\ref{doppler1}) resulted in the gradual increase of the central `hole' of small velocities in our DMs. This central dip involves the additional absorption imposed on the `standard'
double-peaked profile produced by the accretion disc. We have checked this using the BP method without the constraint of a positive
signal in the reconstructed DMs. Indeed, the central hole carried negative signals for all the Balmer lines.What is more, this negative
signal increased from cycle to cycle. It cannot be excluded that the V-shape dip in the emission-line profiles, which produces this
central hole, is generated by the Keplerian shear broadening \citep{horne1995} especially visible for high inclination systems such as V2051 Oph.
This circumstance can distort significantly the prospective measurements of the outer disc radius based on the double-peak separation.

The most impressive feature revealed by our H$\alpha$ maps is a single-armed spiral structure. Such structure is in agreement with the synthetic DMs obtained for another cataclysmic variable star IP~Peg
\citep{kuznetsov2001}. Their three-dimensional simulations show that the region of increased brightness in the H$\alpha$ map can be interpreted as the shock wave along the edge of the gas stream  (the so-called hot line). It originates from the interaction between the gas stream and the circumbinary envelope.

However, almost all of our DMs except for H$\alpha$ and HeI$\lambda$5876 reveal two bright regions in the top right (first) and bottom left (third) quadrants.
Although HeI$\lambda5876$ maps also exhibit similar regions, they are located at different positions.
This difference is likely the result of a nearby strong blend and therefore the DMs of HeI$\lambda5876$ have to be interpreted with caution.
According to theory, such arch-shape structures are created by the tidal influence of the secondary star on the accretion disc. Similarly located structures but in hot discs can generate hydrodynamical shocks, which are able to transport angular momentum.
\cite{steeghs1997} revealed the strong spiral arms in the 1st and 3d quadrants of their maps of IP~Peg observed in active state.

The hot spot is much less pronounced in our maps compared with \citet{longa-pena2015}
and \citep{papadaki2008}.  
 This is likely the result of the significant angular smearing in our DMs introduced by relatively long integration time. In most cases, the
hot spot is practically unnoticeable with the exception of a small bulge located below the ballistic stream trajectory (well visible for
the first cycle of H$\gamma$, HeI $\lambda 7065$ and HeI $\lambda 6678$). In the consecutive
cycles, the central additional absorption (described previously) rises and destroys structures located at the smallest velocities of the
DMs. However, the overall behaviour seems to be clear -- for an increasing wavelength of the subsequent Balmer lines, the regions
of intensified brightness in the DMs move outside (geometrically closer to the white dwarf). 

Another detail well visible in all the maps is a decline of emission intensity in the zone located between the Keplerian and ballistic
trajectories. This is likely to be the result of shadowing of disc emission by gas flowing towards the accretion disc. Another place
of reduced brightness in some DMs (best visible in H$\gamma$, HeI$\lambda$6678 and HeI$\lambda$7065) is located in the outermost part of the disc at the border between the third and fourth quadrants. This might be an effect of the overflowing stream that is re-entering the disc. The simulations carried out by \citet{kunze2001} show such a possibility. 

\begin{figure} 
\includegraphics[width=0.35\textwidth,angle=-90]{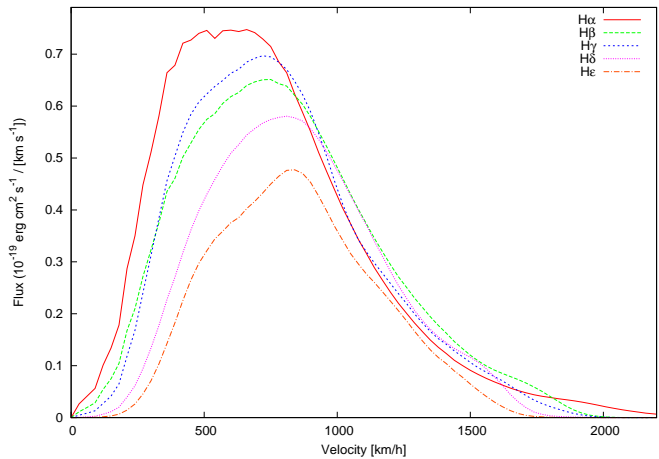}
\caption{Radial profiles inferred from the DMs of five Balmer lines. 
} \label{diskradius}
\end{figure}

We prepared a mean DM for each Balmer line, averaging maps for three consecutive cycles. We then computed the radial profiles
from these mean DMs by averaging the signal in rings for different velocity distances from the white dwarf (see Fig.~\ref{diskradius}). 
We tried to estimate the inner accretion disc radius by measuring the extension of the high-velocity wings of these profiles. Because there are some differences between the high-velocity parts of the profiles for different Balmer lines, we take the mean extent as a good measure of
the inner disc radius. In addition, we exclude the H$\alpha$ profile from averaging as it has quite a different character and its DM shows,
contrary to other lines, one arm spiral. Eventually, we have obtained $V = (1920\pm101)$ km/s.  
Adopting that the radius of the white dwarf is 0.0103 
solar radius \citep{baptista1998} and accepting the Keplerian character of the disc, this velocity corresponds to $3.87^{+0.44}_{-0.39} {\rm R_{wd}}$.

Other researchers also used emission-line profiles to estimate the inner radius of the accretion disc in V2051~Oph. \citet{warner1987} have found $\sim1777$ km/s, which means $\sim$ $4.4\rm{R_{wd}}$, and \citet{watts1986} derived $\sim2000$ km/s which gives $\sim3.6 \rm R_{wd}$.
However, \citet{steeghs2001} found much broader line wings but, as mentioned by \citet{vrielmann2002}, their lines could be enhanced by the disc wind because during the observations V2051~Oph was declining from outburst. The fact that the accretion disc is truncated before it reaches the surface of the white dwarf agrees with the suggestion of \citet{warner1987} that V2051~Oph may be an intermediate polar.
\section{Maps of temperature distribution} \label{temperaturemaps}
\begin{figure} 
\includegraphics[width=0.45\textwidth]{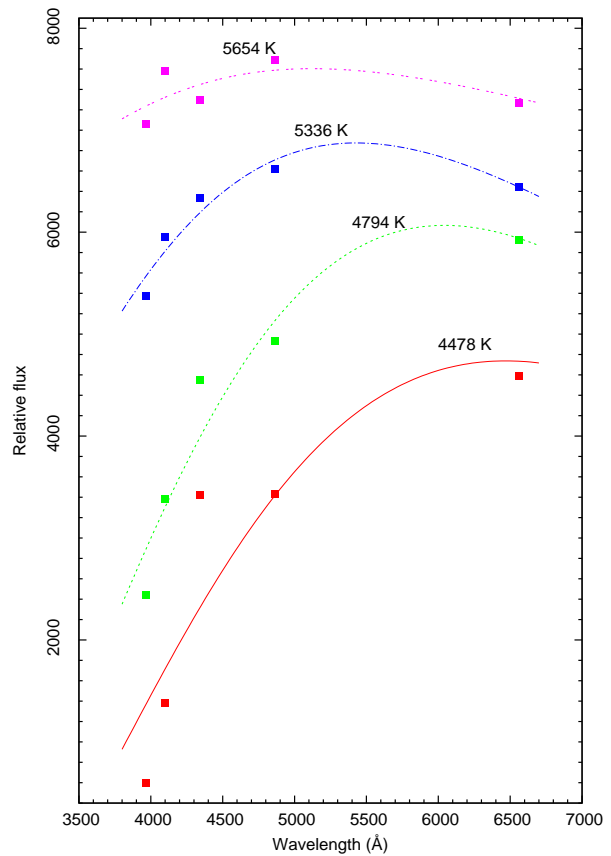}
\caption{Result of fitting the Planck function to the signals from the DMs for the Balmer series. Fits were made at four exemplary points having different velocity distances from the map centre. Profiles were shifted relatively to each other for better visualization.}\label{planck}
\end{figure}
DMs obtained by Doppler tomography reveal the brightness distribution deduced from the intensity profiles of emission lines. This
brightness distribution is a derivative of the physical conditions of the medium in which the considered line is generated and for
a given infinitesimal element in the DM it depends mainly on the gas column density and gas temperature. The standardDMdoes not
showwhether the observed structures are a result of inhomogeneous density or inhomogeneous temperature distribution. Therefore, it
is highly desirable to determine at least one of these parameters separately.
\citet{horne1985} introduced the technique called eclipse mapping. This technique uses information about the accretion disc
hidden in the eclipse light curve of high-inclination systems. As the original method involves the brightness temperature derived
from a single hydrogen emission line, it needs knowledge of the system--observer distance, which could be difficult to obtain with
proper accuracy. Based on this method, \citet{vrielmann1999, vrielmann2002} developed the physical
parameter eclipse mapping (PPEM) approach and decoupled the temperature and surface density of V2051 Oph. Our approach to revealing
the temperature distribution fromDoppler tomography is, to our knowledge, the first attempt of this kind. We took advantage of
having sufficiently strong and uncontaminated lines of the Balmer series. The fundamental assumption of our approach is that the disc
can be represented by an isothermal, isobaric pure-hydrogen slab and the intensity of light emitted from its surface by each infinitesimal
region can be computed from the formula \citep{gaensicke1999}:
\begin{equation}
I_{\lambda}(T)=B_{\lambda}(T)\left[1-e^{-\tau_{\lambda}}\right] \label{eq:intensity}
\end{equation}
Here, $B_{\lambda}(T)$ is the Planck function, $\tau_{\lambda}$ is the wavelengthdependent
optical depth, which involves gas column density and the mass absorption coefficient. The above model is able to give
quite a realistic synthetic spectra of the disc containing systems. At this point, we take a simplistic assumption that the optical thickness
in the centres of the emission lines $\tau_{\lambda}>1$ and is markedly higher than in the continuum. If so, intensity decrements between
consecutive Balmer lines should be described by a blackbody distribution better and better as the optical depth increases. Our bold
assumption about the thermal origin of the Balmer series in the disc, which is optically thick for this series, is in agreement with
the models of \citet{williams1980} and \citet{tylenda1981} showing that peak fluxes of the Balmer lines in cataclysmic variables should fall
on the Planck curve. We are aware of the shortcomings in our approach. However, if the above condition regarding sufficiently large
optical thickness is not fulfilled, one could use an adequate model to compute the realistic extinction coefficient in the accretion disc. Such a model should include at least bound-bound, bound-free and free-free H transitions and H$^{-}$ absorption \citep[see e.g.][]{gaensicke1997,gaensicke1999}. 
The last assumption of our method is that the superposition of all line broadenings such as natural, thermal, collisional, turbulent and instrumental is well below the pixel size of the DMs (30 km s$^{-1}$). Taking into account the overall physical parameters of the quiescent disc we are rather sure this condition is fulfilled.  

Having Doppler maps for five Balmer emission lines (form H$\alpha$ to H$\epsilon$) we can try to obtain the map of temperature distribution in velocity space (as a standard DM). To do this, we have fitted the Planck function to five data points, each placed at the same velocity position in the DMs for subsequent Balmer lines. Fig.~\ref{planck} shows how our approach and accepted assumptions work in practice.We chose
four different points at different distances from the centre of the DM to illustrate how well the Planck function describes the dependency
between flux and wavelength. It can be seen that fits are quite good, which seems to confirm our assumption about the optically
thick accretion disc for the Balmer lines. Of course, one should remember that equation (4), on which our approach is based, could
not be a valid description of the Balmer line emission given the presence of absorption components and saturation effects. In addition,
we have not taken into account the saturation broadening and Keplerian shear broadening \citep{horne1995} that can easily exceed the 30~km~s$^{-1}$ limit (one elementary domain in our DMs) affecting the line emission and hence biasing our temperature maps. One
should also bear in mind that our method can probe the temperature in the region where the emission lines are produced, the region
outside the mid-plane of the disc and where the temperature inversion can occur. Taking into account all the objections regarding our
method of retrieving the temperature maps, they should be treated with caution and only as a crude approximation of the real physical conditions in the accretion disc. It is also obvious that our method can give adequate temperature only at the points in the DM
where the unequivocal relation between velocity position and geometrical position exists. This ensures that the infinitesimal region
in velocity space for which the temperature is derived from Planck fitting corresponds to the infinitesimal fraction of the space map.
Such a situation takes place for the disc with the unequivocal function connecting velocity and distance from the centre (e.g.Keplerian
disc). In contrast, the parts of the velocity space in DMs occupied by the overlapping accretion disc and the secondary star or by the gas
stream do not fulfil the above criterion and the obtained temperature could be erroneous. Thus, we avoided such a situation by choosing
four points, for which the exemplary fits are presented in Fig.~\ref{planck}. Analyzing this figure, one could get the impression that our Planck fitting procedure is completely controlled by the H$\alpha$ line. To check this, we have obtained  temperature maps with the H$\alpha$ DM rejected from computations. The differences between both approaches appeared to be at the level of a few hundred Kelvins. However, the lack of the very important H$\alpha$ line resulted in much higher temperature errors yet rarely exceeding 10\%. Furthermore, we note that the procedure of fitting the Planck function for the Balmer series works properly over a limited T range (from a few thousands to a couple of tens of thousand K).

\begin{figure*} 
\includegraphics[width=0.9\textwidth]{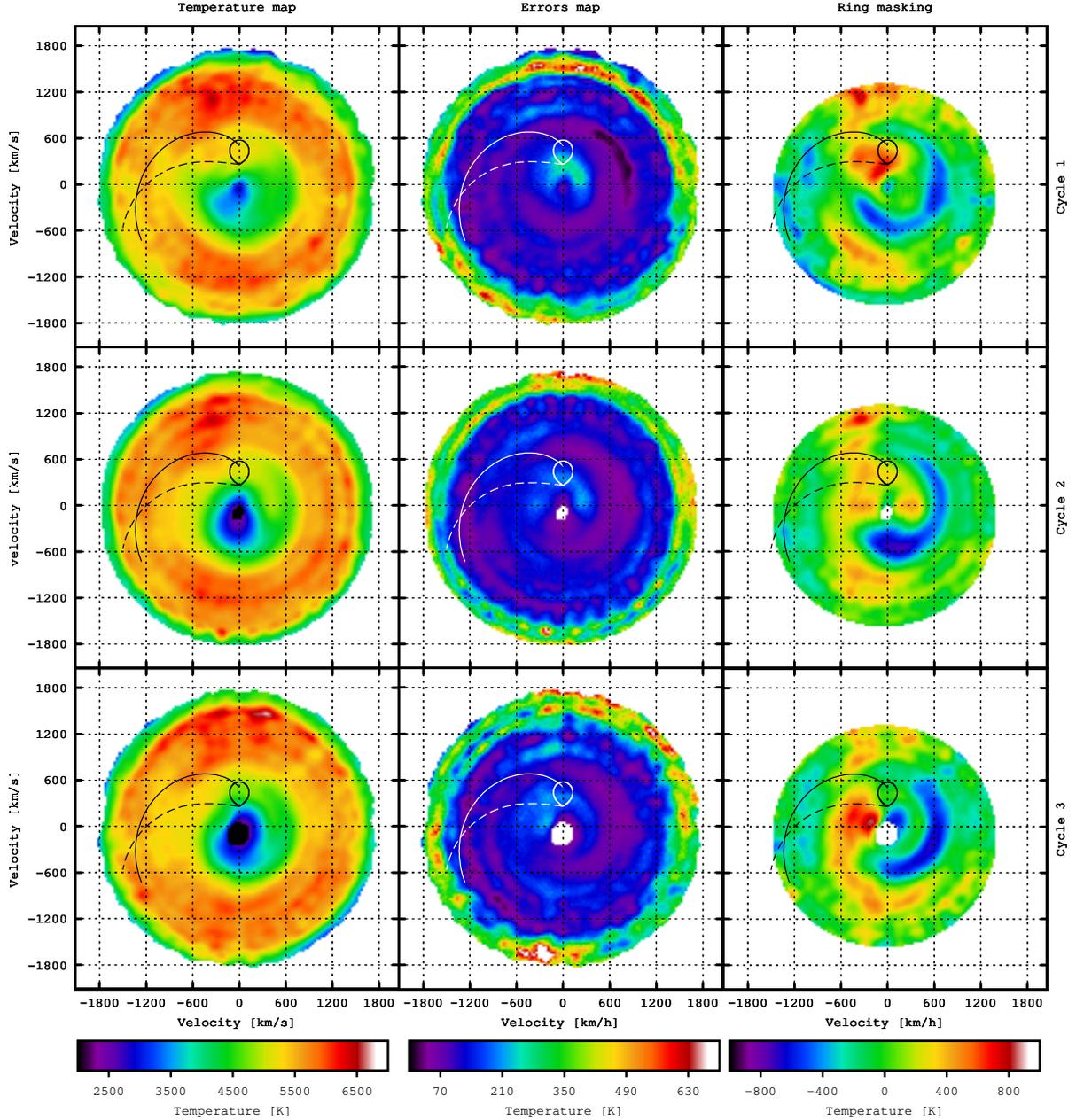}
\caption{Temperature distribution for V2051 Oph in the velocity space. Left column shows temperature maps, the middle column shows temperature errors and the right column shows temperature maps after the `ring masking' enhancement.  At the bottom of each column, the colour temperature scale is included.
}\label{temperature}
\end{figure*}

Our temperature maps and corresponding error maps for the three orbital periods are presented in Fig.~\ref{temperature}. 
The error maps have been computed using the Monte Carlo approach. For each pixel, we have determined the mean square discrepancy between intensity points and the Planck function fitted (see Fig.~\ref{planck}). We treated this discrepancy as a dispersion measure of the Gaussian distribution. From this distribution, 500 random generations of the groups of intensity points have been realized and for each generation the
Planck function fit has been performed. Thus, we have obtained 500 temperature values and computed the temperature dispersion.
Unfortunately, taking into account our simplistic assumptions that are behind our approach as well as many effects that can bias our
temperature maps, most probably systematic errors are dominant here. A surprising conclusion can be drawn that strong spiral such
as structures visible in our DMs do not produce any regions of enhanced temperature in the temperature maps. The problem can
be resolved if one takes into account the properties of the Planck function. The features in our DMs that have a contrast of a few tens
of per cent indicate a couple of per cent contrast of temperature gain. One should also bear in mind that the transformation of the
surface luminosity distribution between $XY$ and $Vx$ $Vy$ spaces for Keplerian disc rotation enhances the visibility of ring- and spirallike
features in the velocity domain.We have performed appropriate tests confirming that the structures visible in our DMs can imply
features with increased temperature, which are almost invisible at the level of error bars and artefacts.

Right column of Fig.~\ref{temperature} shows the result of the so-called ring masking procedure implemented for our temperature maps.
This method subtracts from the signal of a given map pixel, the average signal computed in the ring centred on the position of the primary component ($0,-K1$). 
We covered the whole map by a set of such concentric rings, each having one pixel width and differing by one pixel in radius. Such an approach allows us to improve the visibility of the features lying on the background with a strong radial gradient (e.g. a radial temperature gradient of the disc). When interpreting Fig.~\ref{temperature}, one should remember that there are two regions where our assumptions about optical thickness and hence about thermal emission do not comply. The first region is the central part of each map (below $\sim200$ km s$^{-1}$) and the second is the outermost ring. Both regions are at the edge of the accretion disc where the column density markedly decreases and the optical thickness tends to zero. Moreover, in the case of the central patch we have to deal with the additional absorption, which has been
mentioned earlier in Section~\ref{dopplermaps}. In the outer region of our temperature distribution (innermost region of the accretion disc), the conditions are quite different from the assumed ones, which is manifested in our error maps. This area of the highest velocities is subject to major temperature errors. The temperature distribution at the surface of the accretion disc shows a general increase by 1000 K
from the central to outermost regions in velocity space, which means a decrease from the centre towards the outer rim of the disc in
accordance with the overall tendency observed among other systems and inferred from model computations. The mean temperature of the
disc appeared to be 5600 K. The patchy structure of the temperature maps is undoubtedly an artefact of the DM reconstruction and the
arc features come most probably from our `phase multiplication by interpolation' procedure.

Neither our temperature maps nor the ring masked versions show undeniably traces of the spiral structure conspicuous in the
DMs. In addition to the explanation of this fact given above, one may notice that expectations of the increased temperature in spiral
arms might not be justified in our case. Although the spiral structures we observe are likely tidally induced, it seems that considering
them as `shocks' is premature. Over time (and under appropriate physical conditions) those places of higher density may
evolve into fully developed shocks. There is no such guarantee, of course.

Analysing the maps enhanced by the ring masking procedure, we can easily notice the region of higher temperature in the predicted
gas stream location between ballistic and Keplerian trajectories. Although this area stretches towards the map centre approaching
dangerously the region of unrealistic temperatures, generally it can be treated as certain.
\section{Temperature radial profile} \label{sec:radial}
\begin{figure} 
\includegraphics[width=0.40\textwidth,angle=-90]{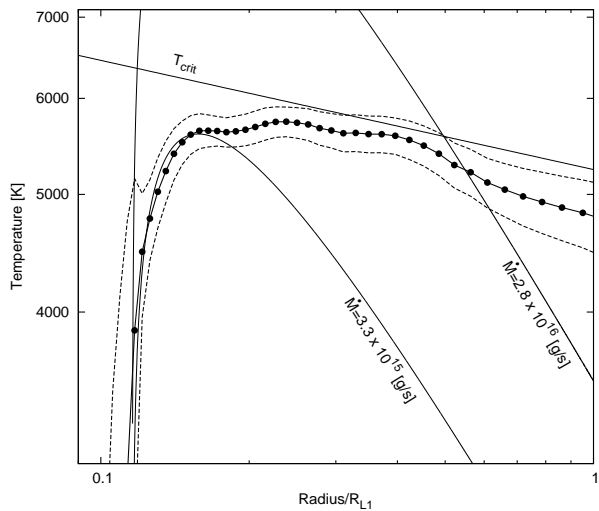}
\caption{Radial temperature of the accretion disc V2051 Oph. Black dots present temperatures determined as mean values in rings of one pixel width and dashed lines show temperature dispersions in these rings. Stationary
accretion disc models with denoted accretion rates are shown by solid curves. The effective critical temperature $T_{\rm crit}$ deduced on the basis of DIM \citep{lasota2001} is shown by a straight line.} \label{radialprof}
\end{figure}
The radial profile of the effective temperature of the stationary accretion disc can be described by the following equation \citep{frank2002}:
\begin{equation}
T_{eff}^4= \frac{3 G M \dot{M}}{8 \pi \sigma R^3}\left(1-\sqrt{R_{\rm w}/R}\right)
\label{eq_temp}
\end{equation}
Here, $G$ and $\sigma$ are gravitational and Stefan-Boltzmann's constants, $R_{\rm w}$ the inner radius of the accretion disc, $M$ is the mass of the accreting object, $\dot{M}$ is the mass accretion rate and $R$ is the radial distance of a given point.
In order to compare our results with the theoretical predictions, we derived the radial temperature profile of
the disc from our temperature maps. First, we averaged three maps for three subsequent orbital cycles and obtained the map with mean
temperatures. Doing this, we neglected the fact that there is some overlapping of the cycles. Next, we computed the radial temperature
profile in the velocity domain averaging data in rings of radii changing by one pixel. Moreover, we obtained temperature dispersions
in these rings. Then, assuming Keplerian velocities in the disc, we recalculated the temperature profile from the velocity space to
the spatial coordinates.
Fig.~\ref{radialprof} presents the result.
In this figure, we have shown the theoretical profiles calculated with equation \ref{eq_temp} in case when $M_1=0.78M_{\odot}$ and the inner disc radius $\sim4.8\pm0.5 R_{\rm wd}$ (estimated   
 by the abrupt decrease in temperature profile visible in the Fig.~\ref{radialprof}).
We can see that this estimation is in general agreement with the estimation of the inner disc radius shown in Section~\ref{dopplermaps}.
The disc instability model (DIM) predicts the existence of the critical effective temperature ($T_{\rm crit}$) defining the state of the accretion disc. 
If the temperature is lower than $T_{\rm crit}$, then the gas in the disc is in neutral, inviscid conditions, which means a quiescent state. \citet{lasota2001} provided a simple equation defining the relation of critical temperature versus disc radius. We show this profile in Fig.~\ref{radialprof}. It is obvious that the system is in quiescence, below the critical temperature level. Our temperature profile is significantly different from the profiles expected for a stationary accretion disc but is qualitatively similar to the profiles obtained with the eclipse mapping technique for this system \citep{baptista2007} and other dwarf novae \citep[e.g.][]{wood1986,wood1989}. Ignoring the rapid fall in the innermost part of the disc, the temperature slowly decreases towards outer regions with values around 5600 K in the intermediate parts. For radii $R < 0.15 R_{L1}$ the temperature drops abruptly, and therefore one can expect a significant change in the physical properties of the medium. This would be consistent with the situation where the inner disc does not reach the surface of the white dwarf. In the region between $R>0.5R_{L1}$ and $R<0.6R_{L1}$ we also see a slightly steeper gradient in the temperature profile, which resembles the $T\propto R^{-3/4}$ law that one can use to estimate the accretion rate. However, the estimated temperature profile does not follow the temperature profile characteristic for a stationary accretion disc in outburst. Instead, it is mostly flat, which is consistent for a quiescent disc. Therefore, we should not use equation \ref{eq_temp} to unambiguously find the accretion rate and we only plotted the theoretical profiles defined by estimated limits of the accretion rates. 
\section{Conclusions} \label{conclusions}
We have presented analysis of the spectral observations of V2051~Oph. Using Doppler tomography, we discovered two-arm spiral
asymmetries in its accretion disc. Amateur photometric observations, overall spectral characteristics and estimations of the accretion
disc temperature show that the system was still in quiescence during our observations, about two weeks before the nearest outburst,
as shown by AAVSO data. Spiral structures are clearly visible in all of the ten studied emission lines. Spiral structures are located
in the first and third quadrants of our maps, which is in agreement with a tidal interpretation of their origin. For the first time, we have
presented the temperature distribution over the accretion disc surface obtained from the DMs for the Balmer emission lines. Using
the Monte Carlo method, we calculated the formal error of this temperature estimation to be lower than 
$\sim 400K$. Our maps reveal a flat distribution of the temperature over the accretion disc. Again, this is characteristic for a quiescent disc. Our brightness DMs show very weak signs of the hot spot but ring-masked temperature maps show the blurred temperature enhancement in the expected location of the hot spot. We found the average accretion disc temperature to be around 5600 K. This shows that the temperature of the accretion disc of V2051 was below the critical limit deduced from the DIM, proving that the disc was in quiescence. Because we do not see an increase of the temperature on the location of the two-arm spiral, the detected structures are likely the result of increased density. At this stage, these structures cannot be considered as strong hydrodynamical shocks that should have left an imprint in our temperature
maps. 
  
We also detected a one-arm spiral in the $H\alpha$ DMs. This spiral starts in the vicinity of the $L_1$ Lagrangian point, then is tightly wound continuing its route over 360 deg. Between Keplerian and ballistic trajectories, it reduces its brightness but continues the spiral path further, reaching $V_X\sim0$ km/s. Three-dimensional modeling of the gas flow obtained by \citet{bisikalo1997,bisikalo1998} reveals a so-called `hot line', which partially resembles the structure that we detected.  
Estimation of the inner radius of the accretion disc from high velocity wings of Balmer lines gave $\sim3.87 R_{\rm wd}$. This supports the suggestion by \cite{warner1987} that the V2051~Oph would be an intermediate polar.

Finally, we would like to  warn the reader, as already done in Section~\ref{temperaturemaps}, to be reasonably sceptical regarding our determination of the disc temperature distribution. We have performed numerous tests proving that our method works well under the indicated assumptions. However, some reservations still exist. Therefore, our methodology and the results obtained require more work and more
and better data for further tests.

\section*{Acknowledgments}
The authors would like to thank the anonymous referee for valuable comments which helped to improve the
manuscript.  Maciej Bilicki is warmly acknowledged for careful reading of the manuscript.
AR is supported by Polish  National  Science  Centre via grant  DEC-2012/04/S/ST9/00021.
This work has partly used the observations provided by the European Organisation for Astronomical Research in the Southern Hemisphere under ESO programme 091.B-0289(F)  and the International Variable Star Index (VSX) database, operated at AAVSO, Cambridge, Massachusetts, USA.











\bsp	
\label{lastpage}
\end{document}